\documentclass[a4paper]{article}

\usepackage{INTERSPEECH2021}
\usepackage{multirow,hyperref}

\DeclareMathOperator*{\relu}{ReLU}
\DeclareMathOperator*{\tanhshrink}{tanhshrink}
\DeclareMathOperator*{\softmax}{softmax}
\DeclareMathOperator*{\stft}{STFT}

\title{High-fidelity and low-latency universal neural vocoder based on multiband WaveRNN with data-driven
linear prediction for discrete waveform modeling}
\name{Patrick Lumban Tobing$^1$, Tomoki Toda$^1$}
\address{
  $^1$Nagoya University, Japan}
\email{patrick.lumbantobing@g.sp.m.is.nagoya-u.ac.jp, tomoki@icts.nagoya-u.ac.jp}

\begin{document}

\maketitle
\begin{abstract}
  This paper presents a novel high-fidelity and low-latency universal neural vocoder framework based on
  multiband WaveRNN with data-driven linear prediction for discrete waveform modeling (MWDLP). MWDLP employs a
  coarse-fine bit WaveRNN architecture for 10-bit mu-law waveform modeling. A sparse gated recurrent unit with a
  relatively large size of hidden units is utilized, while the multiband modeling is deployed to achieve real-time
  low-latency usage. A novel technique for data-driven linear prediction (LP) with discrete waveform modeling
  is proposed, where the LP coefficients are estimated in a data-driven manner. Moreover, a novel loss function
  using short-time Fourier transform (STFT) for discrete waveform modeling with Gumbel approximation is also
  proposed. The experimental results demonstrate that the proposed MWDLP framework generates high-fidelity
  synthetic speech for seen and unseen speakers and/or language on 300 speakers training data including clean
  and noisy/reverberant conditions, where the number of training utterances is limited to 60 per speaker, while
  allowing for real-time low-latency processing using a single core of $\sim\!$~2.1--2.7~GHz CPU with
  $\sim\!$~0.57--0.64 real-time factor including input/output and feature extraction.
\end{abstract}

\section{Introduction}

A neural vocoder \cite{Tamamori17,Ai18,Oord18} utilizes a neural network model to synthesize speech waveform
samples from higher-level input conditioning, e.g., spectral-harmonic features. The use of neural vocoder has
been a common feat in speech synthesis topics in recent years, surpassing the usage and
the performance \cite{Shen18, Zhao20} of conventional vocoders \cite{Kawahara99,Morise16}. In practice,
there exists different types of neural vocoder architecture, which will be more suitable for one use case
than another. Hence, it is worthwhile to develop a strong basis framework that can be flexibly deployed
with meticulous requirements, such as high-fidelity output, real-time low-latency processing with
low-computational machine, and multispeaker training data.
    
Generally, neural vocoder architectures can be categorized into two: autoregressive (AR)
\cite{Oord16,Kalchbrenner18,Jin18,Valin19} and non-AR \cite{Wang18,Prenger19,Kumar19,Yamamoto20}, where the
former is based on sample-dependent synthesis and the latter is based on sample-independent synthesis.
In practice, it is more difficult to optimize non-AR models for low-latency real-time processing while
maintaining the performance due to the usual utilization of multiple layers (deep) of convolutional network.
In this work, to handle low-latency usage in a more straightforward manner, a compact and sparse AR model based
on recurrent neural network (WaveRNN) \cite{Kalchbrenner18,Valin19} is utilized, where sequential computation,
as in a low-delay streaming application, instead of parallel computation can still be achieved in real-time.

Essentially, the quality of a compact and sparse WaveRNN will be more limited compared to a larger and/or dense
model \cite{Kalchbrenner18,Valin19}. Therefore, it is necessary to increase the model capacity (hidden units),
while still considering the size of the model footprint. As increasing hidden units also adds more
complexity, multiband modeling \cite{Okamoto18,Yu19,Tian20} can be used to reduce the complexity for real-time
low-latency applications. Henceforth, in this work, we utilize the use of multiband modeling for a sparse
WaveRNN that employs relatively large hidden units for the gated recurrent unit (GRU) \cite{Cho14}.

\begin{figure*}[t!]
  \centering
  \includegraphics[width=0.72\textwidth]{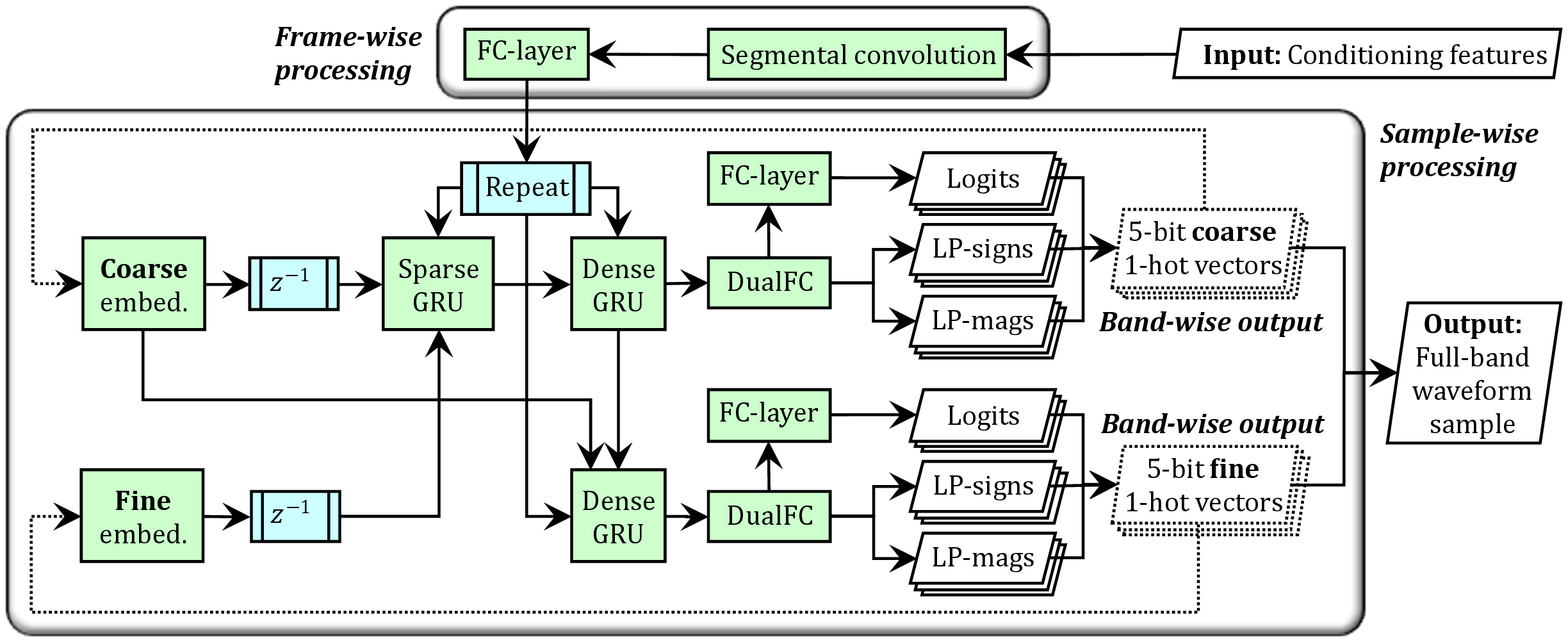}
  \vspace*{-2.5mm}
  \caption{Diagram of the proposed multiband sparse WaveRNN with data-driven linear prediction (MWDLP)
  using 10-bit mu-law output architecture with coarse and fine bits.}
  \label{fig:system_flow}
  \vspace*{-5mm}
\end{figure*}

Lastly, to enhance the model capability of handling multispeaker data (universal neural vocoder) as
well as of producing high-fidelity output, we propose two novel techniques for discrete waveform modeling. First,
we propose to use a data-driven linear prediction (LP) \cite{Atal71} technique for discrete waveform modeling,
where the LP coefficients are estimated in a data-driven manner. Secondly, we propose to use loss
function based on short-time Fourier transform (STFT) \cite{Yamamoto20} with Gumbel approximation \cite{Maddison14} for discrete
waveform modeling. These proposed methods are applied on a sparse multiband WaveRNN that utilizes coarse-fine bit
architecture for discrete modeling of 10-bit mu-law \cite{itut88} waveform, which is called multiband
WaveRNN with data-driven linear prediction (MWDLP). The experimental results demonstrate that the proposed MWDLP
is able to generate high-fidelity synthetic speech for seen and unseen conditions with 300 speakers training
data, where each speaker is limited to 60 training utterances, while allowing real-time low-latency usage on
low-computational machines, which is, to the best of our knowledge, has never been achieved before.




\section{Proposed MWDLP framework for discrete waveform modeling}

Let $\vec{s}=[s_1,\dotsc,s_{t_s},\dotsc,s_{T_s}]^{\top}$ be the sequence of discrete waveform samples and
$\vec{c}=[\vec{c}^{\top}_1,\dotsc,\vec{c}^{\top}_{t_f},\dotsc,\vec{c}^{\top}_{T_f}]^{\top}$ be the sequence of
conditioning feature vectors, where $\vec{c}_{t_f}$ is a $d$-dimensional input feature vector. The
sample-level sequence length is denoted as $T_s$ and that of frame-level is denoted as $T_f$. Consider that the
number of bands in multiband processing as $M$, then, the sequence of waveform samples for the $m$th-band is
denoted as $\vec{s}^{(m)}=[s^{(m)}_{1},\dotsc,s^{(m)}_{t},\dotsc,s^{(m)}_{T_m}]^{\top}$, where the length of
subband waveform is denoted as $T_m = T_s / M$. Hence, the sequence of upsampled (repeated)
conditioning feature vectors is denoted as
$\vec{c}^{(u)}=[\vec{c}^{(u)^{\top}}_1,\dotsc,\vec{c}^{(u)^{\top}}_t,\dotsc, \vec{c}^{(u)^{\top}}_{T_m}]^{\top}$.
The objective is to model the probability mass function (p.m.f.) of the discrete waveform as
\begin{equation}
    p(\vec{s})=\!\prod_{m=1}^{M}\prod_{t=1}^{T_m}p(s^{(m)}_t|\vec{c}^{(u)}_t\!,\vec{s}^{(M)}_{t-1})
        =\!\prod_{m=1}^{M}\prod_{t=1}^{T_m}\vec{p}^{(m)^{\top}}_t\!\!\!\vec{v}^{(m)}_t,\!\!
\label{eq:pdf}
\end{equation}
where $\vec{s}^{(M)}_{t-1}=[s^{(1)}_{t-1},\dotsc,s^{(m)}_{t-1},\dotsc,s^{(M)}_{t-1}]^{\top}$,
$\vec{p}^{(m)}_t\!\!=\![p^{(m)}_t[1],\dotsc,p^{(m)}_t[b],\dotsc,p^{(m)}_t[B]]^{\top}\!\!\!$,
$\vec{v}^{(m)}_t\!\!=\![v^{(m)}_t[1],\dotsc,v^{(m)}_t[b],\dotsc,v^{(m)}_t[B]]^{\top}\!\!\!$,
$\sum_{b=1}^{B}v^{(m)}_t[b]\!=\!1$, $v^{(m)}_t[b]\!\in\!\{0,\!1\}$, $B$ is the number of sample bins,
$\vec{v}^{(m)}_t$ is a 1-hot vector, and $\vec{p}^{(m)}_t$ is a probability vector (network output).

\subsection{Data-driven LP for discrete modeling}
\label{ssec:ddlpc}
\vspace{-1mm}

In this work, we propose to use a data-driven LP \cite{Atal71} technique to compute the
probability vector of discrete sample bins $\vec{p}^{(m)}_t$ in Eq.~\eqref{eq:pdf}. Specifically, the probability
of each sample bin $p^{(m)}_t[b]$ is given by the $\softmax$ function as follows:
\begin{equation}
    p^{(m)}_t[b] = \frac{\exp(\hat{o}^{(m)}_t[b])}
            {\sum_{j=1}^{B}\exp(\hat{o}^{(_m)}_t[j])},
\label{eq:prob}
\end{equation}
where $\exp(\cdot)$ denotes the exponential function, $\hat{o}^{(m)}_t[b]$
is the unnormalized probability (logit) of the $b$th sample bin for the $m$th band
at time $t$, and the vector of logits containing all sample bins is given as
$\hat{\vec{o}}^{(m)}_t=[\hat{o}^{(m)}_t[1],\dotsc,\hat{o}^{(m)}_t[b],\dotsc,\hat{o}^{(m)}_t[B]]^{\top}$.

Then, the proposed data-driven LP for discrete waveform modeling is formulated as follows:
\begin{equation}
\hat{\vec{o}}^{(m)}_t = \sum_{k=1}^{K}a^{(m)}_t[k]\vec{v}^{(m)}_{t-k} + \vec{o}^{(m)}_t,
\label{eq:ddlp}
\end{equation}
where the residual logit vector is denoted as $\vec{o}^{(m)}_t$, the $k$th data-driven LP coefficient of
the $m$th band at time $t$ is denoted as $a^{(m)}_t[k]$, $k$ denotes the index of LP coefficient, and the
total number of coefficients is denoted as $K$. The data-driven LP coefficient vector containing all
coefficients for the $m$th band at time $t$ is given as
$\vec{a}^{(m)}_t=[a^{(m)}_t[1],\dotsc,a^{(m)}_t[k],\dotsc,a^{(m)}_t[K]]^{\top}$.
In Eq.~\eqref{eq:ddlp}, $\{\vec{v}^{(m)}_{t-1},\dotsc,\vec{v}^{(m)}_{t-K}\}$ are used as logit basis vectors
corresponding to past $K$ discrete waveform samples $\{s_{t-1},\dotsc,s_{t-K}\}$, which are used for LP
in the logit space.

\subsection{Network architecture}
\label{ssec:nnet}
\vspace{-1mm}

The network diagram of the proposed MWDLP framework for the modeling of 10-bit mu-law waveform is depicted in
Fig.\ref{fig:system_flow}. Conditioning input features are fed into a segmental convolution layer that takes into
account $r=5$ previous and $n=1$ succeeding frames to produce a $((r+1+n)\times d)$-dimensional feature vector
from $d$-dimensional input feature vectors, which is then passed to a fully connected (FC) layer with
$320$-dimensional output and $\relu$ activation. Separate embedding layers with $64$-dimensionality are used to
encode 1-hot vectors of 5-bit fine- and 5-bit coarse-parts of the waveform sample, respectively, which are shared
between all bands. Sparse GRU has a relatively large number of hidden units ($1184$), while two separate dense
GRUs have small number of hidden units ($32$).

Separate dual fully-connected (DualFC) layers are used for the fine- and coarse-bit outputs. Each DualFC layer
produces two output channels that are combined by a trainable weighting vector, as in \cite{Valin19},
where the weighting vector is activated by $\exp$ function and multiplied by a constant $0.5$. Each
output channel of the DualFC consists of the parts that correspond to the data-driven LP vectors
$\vec{a}^{(M)}_t=[\vec{a}^{(1)^{\top}}_t,\dotsc,\vec{a}^{(m)^{\top}}_t,\dotsc,\vec{a}^{(M)^{\top}}_t]^{\top}$
and to the logit vector
$\vec{o}^{(M)}_t=[\vec{o}^{(1)^{\top}}_t,\dotsc,\vec{o}^{(m)^{\top}}_t,\dotsc,\vec{o}^{(M)^{\top}}_t]^{\top}$.
The output part of the data-driven LP vectors consists of signs (LP-signs)
$\vec{a}^{(sg_M)}_t
    =[\vec{a}^{(sg_1)^{\top}}_t,\dotsc,\vec{a}^{(sg_m)^{\top}}_t,\dotsc,\vec{a}^{(sg_M)^{\top}}_t]^{\top}$,
i.e., with hyperbolic tangent ($\tanh$) activation, and of magnitudes (LP-mags)
$\vec{a}^{(mg_M)}_t
    =[\vec{a}^{(mg_1)^{\top}}_t,\dotsc,\vec{a}^{(mg_m)^{\top}}_t,\dotsc,\vec{a}^{(mg_M)^{\top}}_t]^{\top}$,
i.e., with $\exp$ activation. The data-driven LP coefficient vector is computed as
$\vec{a}^{(M)}_t=\vec{a}^{(sg_M)}_t\odot\vec{a}^{(mg_M)}_t$, where $\odot$ denotes the
Hadamard product. The last FC layers with $16$-dimensionality input (from the DualFC logits-part output)
on $\relu$ activation, $32$-dimensionality output on $\tanhshrink$ activation ($x-\tanh(x)$), and
shared over all bands, produce the residual logit vector $\vec{o}^{(m)}_t$.

\vspace{-2mm}
\subsection{STFT-based loss function for discrete modeling}
\label{ssec:stft_loss}
\vspace{-1.5mm}

In this work, we also propose an additional loss function based on STFT
\cite{Yamamoto20} for discrete waveform modeling, where Gumbel sampling \cite{Maddison14} method is utilized. Specifically,
it is used to obtain a sampled probability vector of each $m$th band at time $t$
$\hat{\vec{p}}^{(m)}_t=[\hat{p}^{(m)}_t[1],\dotsc,\hat{p}^{(m)}_t[b],\dotsc,\hat{p}^{(m)}_t[B]]^{\top}$,
where a sampled probability of each $b$th bin $\hat{p}^{(m)}_t[b]$ is given by
\begin{equation}
   \hat{p}^{(m)}_t[b] = \frac{\exp(\hat{\gamma}^{(m)}_t[b])}
            {\sum_{j=1}^{B}\exp(\hat{\gamma}^{(m)}_t[j])},
\label{eq:sampled_prob}
\end{equation}
In Eq.~\eqref{eq:sampled_prob}, $\hat{\gamma}^{(m)}_t[b]$ denotes a sampled logit, where
a sampled logit vector
$\hat{\vec{\gamma}}^{(m)}_t
    =[\hat{\gamma}^{(m)}_t[1],\dotsc,\hat{\gamma}^{(m)}_t[b],\dotsc,\hat{\gamma}^{(m)}_t[B]]^{\top}$
is computed as
\begin{equation}
\hat{\vec{\gamma}}^{(m)}_t = \hat{\vec{o}}^{(m)}_t - \log(-\log(\vec{u})), \text{s. t.}\:\vec{u}\sim(0,1),
\label{eq:gumbel_sampling}
\end{equation}
and $\vec{u}$ is a uniformly distributed $B$-dimensional vector.

Then, the discrete value of the sampled waveform bin $\hat{s}^{(m)}_t$ can be recovered while keeping
the backpropagation path from the reparameterization with Gumbel sampling in Eq.~\eqref{eq:gumbel_sampling} as
\begin{align}
&\hat{s}^{(m)}_t= f\bigg(\sum_{b=1}^{B}\:b\:\:\overline{\hat{p}}^{(m)}_t[b]\bigg),\:\:\text{s. t.}\nonumber\\
\overline{\hat{p}}^{(m)}_t[b]&=
\begin{cases}
\frac{\hat{p}^{(m)}_t[b]}{\max{(\hat{\vec{p}}^{(m)}_t)}},
    & \!\!\text{if}\:\:\hat{p}^{(m)}_t[b]=\max{(\hat{\vec{p}}^{(m)}_t)},\\
0, & \!\!\text{else,}
\end{cases}
\label{eq:sampled_values}
\end{align}
where $\max(\vec{p})$ is a function that returns the maximum value of a vector $\vec{p}$ and $f(b)$ denotes
a differentiable function that returns the waveform value of a discrete sample bin $b$, e.g., an inverse mu-law
\cite{itut88} function. Hence, the STFT-based loss is computed from the sampled waveform
$\hat{\vec{s}}^{(m)}=[\hat{s}^{(m)}_1,\dotsc,\hat{s}^{(m)}_t,\dotsc,\hat{s}^{(m)}_{T_m}]^{\top}$
and the target waveform $\vec{s}^{(m)}$ as
\begin{equation}
\mathcal{L}_{\text{STFT}}^{(m)} = g(\stft(\vec{\hat{s}}^{(m)}),\stft(\vec{s}^{(m)})),
\label{eq:stft}
\end{equation}
where $\stft(\cdot)$ denotes an STFT analysis function that produces frames of complex STFT spectra and
$g(\cdot,\cdot)$ denotes a set of STFT-based loss functions. Ultimately, the loss of full-band waveform $\vec{s}$
can also be computed as in \cite{Yang20}. Note that the discretization through thresholding with $\max(\cdot)$
function in Eq.~\eqref{eq:sampled_values} is necessary to accommodate the $f(\cdot)$ function.

\subsection{Sparsification and model complexity}
\label{ssec:sparse_model}

In training, a sparsification procedure is performed for the recurrent matrices of the large sparse $1184$ GRU
in Fig.~\eqref{fig:system_flow}, where the average target density from all recurrent matrices of update,
reset, and new gates \cite{Cho14,Valin19} is $0.1$, and each target densities are $0.09$, $0.09$, and $0.12$,
respectively. The complexity is computed as in \cite{Valin19,Yu19} with adjustments according to the MWDLP
architecture. For a 24 kHz waveform model with $M=6$ bands and $K=8$ LP coefficients, the total complexity of
the band-rate module is $\sim\,\!\!4.53$ GFLOPS, while for a 16 kHz model with $M=4$ and $K=8$, it is
$\sim\,\!\!4.24$ GFLOPS.


\begin{table}[!t]
  \scriptsize
  \caption{Training/Development speech dataset configurations. The number of training utterances per speaker is
  limited to $60$.}
  \vspace{-2.5mm}
  \label{tab:dataset}
  \centering
  \begin{tabular}{c c c}
    \toprule
    \multicolumn{1}{c}{\textbf{Language/dialect/condition}}
        & \multicolumn{1}{c}{\textbf{\# Male}}
            & \multicolumn{1}{c}{\textbf{\# Female}} \\
    \midrule
    Spanish (4 dialects) \cite{Guevara20}             & $20$~~~               & $20$~~~               \\
    Catalan, Galician \cite{Kjartansson20}            & $10$~~~               & $10$~~~               \\
    Yoruba \cite{Gutkin20}, isiXhosa \cite{Niekerk17} & $5$~~~                & $12$~~~              \\
    Gujarati, Marathi \cite{He20}                     & $5$~~~                & $14$~~~              \\
    Tamil, Telugu \cite{He20}                         & $10$~~~               & $10$~~~              \\
    Bengali (Bangladeshi, Indian) \cite{Sodimana18}   & $13$~~~               & $1$~~~              \\
    Javanese, Khmer \cite{Sodimana18}                 & $5$~~~                & $15$~~~              \\
    French (Emotional/Expressive) \cite{Le20}         & $4$~~~                & $4$~~~              \\
    Japanese \cite{Takamichi19}                       & $29$~~~               & $29$~~~              \\
    English \cite{Veaux16}                            & $28$~~~               & $28$~~~              \\
    English (Noisy/Reverberant) \cite{Valentini17}    & $14$~~~               & $14$~~~              \\
    \bottomrule
  \end{tabular}
  \vspace{-2.5mm}
\end{table}

\begin{table}[!t]
  \scriptsize
  \caption{Real-time factor (RTF) of MWDLP with $4$~kHz resolution per band and $8$ data-driven LP coefficients, which includes input/output and feature extraction (I/O + feat.).}
  \vspace{-2.5mm}
  \label{tab:rtf}
  \centering
  \begin{tabular}{c c c}
    \toprule
    \multicolumn{1}{c}{\textbf{RTF w/ I/O + feat. on 1-core CPU}}
        & \multicolumn{1}{c}{\textbf{16 kHz}}
            & \multicolumn{1}{c}{\textbf{24 kHz}} \\
    \midrule
    Intel® Xeon® Gold 6230 $2.1$~GHz                  & $0.58$              & $0.64$~~~               \\
    Intel® Xeon® Gold 6142 $2.6$~GHz                  & $0.57$              & $0.63$~~~              \\
    Intel® Core™ i7-7500U $2.7$~GHz                   & $0.57$              & $0.63$~~~              \\
    \bottomrule
  \end{tabular}
  \vspace{-4.5mm}
\end{table}


\section{Experimental evaluation}

\subsection{Experimental conditions}
\label{ssec:exp_cond}

We used speech data from $300$ speakers
\cite{Guevara20,Kjartansson20,Gutkin20,Niekerk17,He20,Sodimana18,Le20,Takamichi19,Veaux16}
consisting of over $18$ languages/dialects including few expressive speech data and noisy/reverberant speech
data. The number of training utterances per speaker was limited to $60$ and the number of development utterances
per speaker was $5$, which were used for early stopping. The details of the training/development speech dataset
are given in Table~\ref{tab:dataset}. Additionaly, for evaluation on unseen conditions, we also utilized speech
data of another $3$ speakers/languages: a male Basque \cite{Kjartansson20}, a female Malayalam \cite{He20},
and a female Chinese \cite{Zhao20} speakers.

For the proposed MWDLP framework, as the conditioning input feature, we utilized $80$-dimensional
mel-spectrogram, which was extracted from the STFT magnitude spectra. In STFT analysis, the shift length was set
to $10$~ms, the window length was set to $27.5$~ms, and Hanning window was used. For $24$~kHz waveform, the FFT
length was set to $2048$, while for $16$~kHz waveform, it was set to $1024$. The ablation objective evaluation
was performed using the $24$~kHz models.

\begin{table}[!t]
  \scriptsize
  \caption{Objective evaluation results excluding noisy/reverberant data with variations of
  data-driven linear prediction (LP) and the use of STFT loss (STFT).}
  \vspace{-2.5mm}
  \label{tab:obj_res_c}
  \centering
  \begin{tabular}{c c c c c}
    \toprule
    \multicolumn{1}{c}{\textbf{Model}}
        & \multicolumn{1}{c}{\textbf{MCD [dB]}}
            & \multicolumn{1}{c}{\textbf{U/V [\%]}}
                & \multicolumn{1}{c}{\textbf{F0 [Hz]}}
                    & \multicolumn{1}{c}{\textbf{LSD [dB]}} \\
    \midrule
    MWDLP 0LP                           & $2.88$~~~      & $12.17$~~~           & $17.28$~~~     & $4.93$~~~          \\
    MWDLP 0LP+STFT                      & $2.97$~~~      & $12.18$~~~           & $17.67$~~~     & $5.05$~~~               \\
    MWDLP 6LP                           & $2.87$~~~      & $13.17$~~~           & $17.53$~~~     & $4.95$~~~               \\
    MWDLP 6LP+STFT                      & $2.91$~~~      & $13.03$~~~           & $17.51$~~~     & $4.93$~~~          \\
    MWDLP 8LP                           & $2.87$~~~       & $12.33$~~~           & $17.25$~~~    & $4.82$~~~           \\
    \textbf{MWDLP 8LP+STFT}             & $\textbf{2.78}$~~~       & $\textbf{12.10}$~~~           & $\textbf{17.25}$~~~    & $\textbf{4.80}$~~~           \\
    \midrule
    PWG                            & $\textbf{2.88}$~~~       & $\textbf{15.44}$~~~           & $\textbf{20.07}$~~~     & $\textbf{4.49}$~~~          \\
    Fatchord                            & $6.03$~~~       & $27.60$~~~           & $17.22$~~~      & $7.32$~~~         \\
    LPCNet                              & $4.09$~~~       & $13.50$~~~           & $21.82$~~~      & $11.92$~~~         \\
    \bottomrule
  \end{tabular}
  \vspace{-4.5mm}
\end{table}

The hyperparameters of MWDLP were set as in Sections~\ref{ssec:nnet} and \ref{ssec:sparse_model}. In training,
dropout with $0.5$ probability was used after the upsampling (repetition) of conditioning input features.
RAdam \cite{Liu19} algorithm was used for the parameter optimization, where the learning rate was set to
$0.0001$. Weight normalization \cite{Salimans16} was used for convolution and fully-connected layers.
The batch sequence length was set to $6$ frames and the batch size was set to $8$. Using a single NVIDIA RTX
2080Ti, the training time for a $24$~kHz model with a number of bands $M=6$, a number of data-driven LP $K=8$
(Section~\ref{ssec:ddlpc}), and $5$ windowing configurations for the STFT-based loss
(Section~\ref{ssec:stft_loss}) was $\sim\!4.8$ days. On the other hand, the real-time factor (RTF) in synthesis
was $0.57$--$0.64$ including input/output and feature extraction, which was obtained using a single core of
$2.1$--$2.7$~GHz CPU as given in Table~\ref{tab:rtf}. The footprint size of the compiled model, i.e., the
executable, was $16$~MB. The software has been made available
at {\scriptsize{\url{https://github.com/patrickltobing/cyclevae-vc-neuralvoco}}\normalsize.

The number of data-driven LP coefficients $K$ was varied to $\{0,6,8\}$. In \cite{Markel76},
it was recommended to use one coefficient per kHz plus two pairs of coefficients, each for spectral slope and
voice quality, which puts $K=8$ to be the most suitable for a MWDLP model with $4$ kHz
band-waveform resolution. The $5$ windowing configurations for the STFT-based loss were set for each
band-resolution and full-band waveforms. On full-band, the FFT lengths were set to $\{2048,1024,512,256,128\}$
for $24$~kHz and $\{1024,512,256,128,128\}$ for $16$~kHz, while the shift lengths were set to
$\{480,240,120,60,48\}$ for $24$~kHz and to $\{320,160,80,40,32\}$ for $16$~kHz. On band-waveform, the FFT lengths
were set to $\{256,128,64,32,32\}$ and the shift lengths were set to $\{80,40,20,10,8\}$. In all cases, the window
lengths were set to $2.5$ multiple of the shift lengths.

Pseudo-quadratic mirror filter (PQMF) \cite{Nguyen94} was used for the multiband analysis and
synthesis \cite{Yu19}. The Kaiser prototype filter configurations were as follows: the order was set to $410$ for
$24$~kHz or to $274$ for $16$~kHz, the $\beta$ coefficient was set to $43.12126$, and the cutoff ratio was set to
$0.1$ for $24$~kHz or to $0.15$ for $16$~kHz. Pre-emphasis with $\alpha=0.85$ was applied to the full-band
waveform before PQMF analysis.

Lastly, for additional baselines, we also developed $24$~kHz waveform models with a publicly available
WaveRNN implementation {\scriptsize{\url{https://github.com/fatchord/WaveRNN}}\normalsize~(Fatchord) and with Parallel WaveGAN
(PWG) \cite{Yamamoto20}, which is a non-AR neural vocoder, and a $16$~kHz model using LPCNet \cite{Valin19}. The
training sets were the same as for MWDLP, given in Table~\ref{tab:dataset}.

\begin{table}[!t]
  \scriptsize
  \caption{Objective evaluation results on noisy/reverberant data. Number of
  data-driven linear prediction (LP) was varied including the use of STFT loss (STFT).}
  \vspace{-2.5mm}
  \label{tab:obj_res_n}
  \centering
  \begin{tabular}{c c c c c}
    \toprule
    \multicolumn{1}{c}{\textbf{Model}}
        & \multicolumn{1}{c}{\textbf{MCD [dB]}}
            & \multicolumn{1}{c}{\textbf{LSD [dB]}} \\
    \midrule
    MWDLP 0LP                           & $2.90$~~~          & $4.37$~~~          \\
    MWDLP 0LP+STFT                      & $2.48$~~~           & $4.24$~~~               \\
    MWDLP 6LP                           & $2.79$~~~           & $4.88$~~~               \\
    MWDLP 6LP+STFT                      & $2.49$~~~           & $4.60$~~~          \\
    MWDLP 8LP                           & $2.57$~~~          & $4.19$~~~           \\
    \textbf{MWDLP 8LP+STFT}               & $\textbf{2.44}$~~~           & $\textbf{4.04}$~~~           \\
    \midrule
    PWG                                 & $\textbf{2.41}$~~~       & $\textbf{4.16}$~~~          \\
    Fatchord                            & $3.49$~~~       & $4.84$~~~         \\
    LPCNet                              & $3.33$~~~       & $5.40$~~~         \\
    \bottomrule
  \end{tabular}
  \vspace{-4.5mm}
\end{table}

\subsection{Objective evaluation}
\vspace{-1mm}

In the objective evaluation, we measured the mel-cepstral distortion (MCD) \cite{Mashimo01}, unvoiced/voiced
decision error (U/V), root-mean-square error of fundamental frequency (F0), and log spectral distortion
(LSD). On the measurements of MCD, U/V, and F0 accuracies, WORLD \cite{Morise16} was used to extract F0 and
spectral envelope, where $28$-dimensional mel-cepstral coefficients were extracted with $0.466$ frequency warping
for $24$~kHz and $0.41$ for $16$~kHz. For log-spectral distortion, $80$-dimensional mel-spectrogram extracted from
the magnitude spectra as in Section~\ref{ssec:exp_cond} was used. To adjust for the phase differences between
synthesized and target waveforms, dynamic-time-warping was computed with respect to the extracted mel-cepstra.
$9969$ and $9752$ testing utterances were used for evaluation without and with noisy/reverberant speech,
respectively.

The result of objective evaluation without noisy/reverberant speech test set is given in
Table~\ref{tab:obj_res_c}. It can be observed that the use of $8$ data-driven LP provides better accuracies on
all MCD, U/V, F0, and LSD compared to without using data-driven LP and with $6$ data-driven LP. The use
of STFT-based loss further improves the model with $8$ data-driven LP yielding the best accuracies
on MCD, U/V, F0 and LSD with values of $2.78$~dB, $12.10~\!\%$, $17.25$~Hz and $4.80$~dB, respectively. On the
other hand, the result of objective evaluation with noisy/reverberant speech test set is given in
Table~\ref{tab:obj_res_n}, where U/V and F0 measurements were reasonably omitted. In this result, it can be
observed that the use of $8$ data-driven LP also provides better MCD and LSD values compared to without using
data-driven LP or $6$ data-driven LP, while the use of STFT-based loss further improves it to yield the best
MCD and LSD with values of $2.44$~dB and $4.04$~dB, respectively. Lower values on noisy/reverberant test set are
mainly due to the non-existence of silent speech regions, especially for LPCNet model, where it tends to produce
unclear/noisy sounds and for Fatchord model, where it generates too much noise/artifact even for clean speech.
Overall, it has been shown that the tendency of consistent improvements is obtained by the proposed
MWDLP with $8$ data-driven LP using STFT-based loss (MWDLP 8LP+STFT). Our preliminary testing by listening on the
speech samples also suggests that the MWDLP 8LP+STFT provides the highest speech quality.

\begin{table}[!t]
  \scriptsize
  \caption{Subjective evaluation results showing mean opinion score (MOS) from seen and unseen sets.}
  \vspace{-2.5mm}
  \label{tab:subj_res}
  \centering
  \begin{tabular}{c c c c}
    \toprule
    \multicolumn{1}{c}{\textbf{Model -- MOS}}
        & \multicolumn{1}{c}{\textbf{Seen}}
            & \multicolumn{1}{c}{\textbf{Unseen}} 
            & \multicolumn{1}{c}{\textbf{All}} \\
    \midrule    
    Original $24$~kHz                   & $4.56\pm0.07$ & $4.57\pm0.07$ & $4.57\pm0.05$           \\
    Original $16$~kHz                   & $4.47\pm0.09$ & $4.56\pm0.09$ & $4.52\pm0.06$           \\
    \textbf{MWDLP \bm{$24$}~kHz}        & $\bm{4.15\pm0.09}$ & $\bm{4.29\pm0.09}$ & $\bm{4.22\pm0.06}$               \\
    \textbf{MWDLP \bm{$16$}~kHz}        & $\bm{3.98\pm0.09}$ & $\bm{4.29\pm0.09}$ & $\bm{4.13\pm0.06}$               \\
    PWG $24$~kHz                        & $3.93\pm0.11$ & $4.20\pm0.10$ & $4.07\pm0.07$           \\
    Fatchord $24$~kHz                   & $2.11\pm0.11$ & $2.13\pm0.13$ & $2.12\pm0.08$          \\
    LPCNet $16$~kHz                     & $3.14\pm0.11$ & $3.22\pm0.10$ & $3.18\pm0.08$           \\    
    \bottomrule
  \end{tabular}
  \vspace{-4.5mm}
\end{table}

\subsection{Subjective evaluation}
\vspace{-1mm}

In the subjective evaluation, we chose the MWDLP 8LP+STFT configurations for the $16$ and $24$~kHz waveform
models, which were also compared with the Fatchord $24$~kHz and LPCNet $16$~kHz models, as well as the original
$16$~ and $24$~kHz waveforms. The number of evaluated speakers from the seen dataset of Table~\ref{tab:dataset}
was 3, which were a Spanish Female (Argentinian) \cite{Guevara20}, a Yoruba male \cite{Gutkin20}, and an English
male \cite{Veaux16} speakers. The number of evaluated unseen speakers/language was $3$ as given in
Section~\ref{ssec:exp_cond}. The number of testing utterances per speaker was $10$, i.e., a total of $60$
listening web-pages. Corresponding languages of the audios were also shown to the listeners. The number of
crowd-sourced listeners from Amazon Mechanical Turk was $20$.

The subjective evaluation result is shown in Table~\ref{tab:subj_res}, where the $5$-scaled mean opinion score
(MOS) values on the speech quality ranging from $1$ (very bad) to $5$ (very good) are given. It can be
clearly observed that the proposed MWDLP gives the best performances compared to the $24$~kHz and $16$~kHz 
baseline models by achieving MOS values of $4.15$ and $4.29$ on seen and unseen data, respectively,
for $24$~kHz model and of $3.98$ and $4.29$ on seen and unseen data, respectively, for $16$~kHz model. It
can also be observed that the scores of seen speakers are lower than unseen speakers, which is due to the better
recording quality for the evaluated unseen speakers. Note that the only systems that can be run real-time with
low-latency processing on CPU are the proposed MWDLP and LPCNet \cite{Valin19}. Samples and demo are available
at {\scriptsize{\url{https://demo-mwdlp-interspeech2021.audioeval.net}}\normalsize.

\vspace{-0.5mm}
\section{Conclusions}
\vspace{-1mm}

We have presented a novel real-time low-latency universal neural vocoder with high-fidelity output based
on multiband WaveRNN using data-driven linear prediction for discrete waveform modeling (MWDLP). The proposed
MWDLP framework utilizes a relatively large number of hidden units for the main RNN module, where sparsification
and multiband modeling approaches are applied to reduce the effective model size and the model complexity.
A novel data-driven linear prediction (LP) technique is proposed for the use in discrete waveform modeling,
where the LP coefficients are estimated in a data-driven manner. Further, a novel approach for short-time
Fourier transform (STFT)-based loss computation on discrete modeling with Gumbel approximation is also proposed.
The results have demonstrated that the MWDLP framework is able to generate high-fidelity synthetic speech,
where it is trained with a $300$ speakers dataset, with $0.57$--$0.64$ real-time factor using a single-core
of $2.1$--$2.7$~GHz CPU.

\vspace{-0.5mm}
\section{Acknowledgements}
\vspace{-1mm}

This work was partly supported by JSPS KAKENHI Grant Number 17H06101 and JST, CREST Grant Number JPMJCR19A3.


\bibliographystyle{IEEEtran}

\bibliography{mwdlp}


\end{document}